\documentclass[conference]{IEEEtran}

\usepackage{latexsym}
\usepackage{amssymb}
\usepackage{amsmath}
\usepackage{epsfig}
\usepackage{graphicx}
\usepackage{psfrag}
\usepackage{url}
\usepackage{algorithm}
\usepackage{algpseudocode}
\usepackage{ifthen}
\usepackage{tikz, pgf,pgffor}
\usepackage{cite}

\renewcommand{\IEEEQED}{\IEEEQEDopen}

\newtheorem{theorem}{Theorem}[section]
\newtheorem{lemma}[theorem]{Lemma}

\newcommand{\MD}{\textit{hash}}

\newcommand{\blockchain}{blockchain}
\newcommand{\Blockchain}{Blockchain}
\newcommand{\blockchains}{blockchains}

\newcommand{\ieeeexplore}{0}

\IEEEoverridecommandlockouts

\graphicspath{{./FIGURES/}}

\title{Bitcoin \Blockchain\ Dynamics: the Selfish-Mine Strategy in the
Presence of Propagation Delay}

\author{%
\IEEEauthorblockN{%
J. G\"{o}bel%
\IEEEauthorrefmark{1},
H.P. Keeler%
\IEEEauthorrefmark{2},
A.E. Krzesinski%
\IEEEauthorrefmark{3}
and P.G. Taylor%
\IEEEauthorrefmark{2}
}%
\IEEEauthorblockA{%
\IEEEauthorrefmark{1}%
Department of Informatics,
University of Hamburg,
22527~Hamburg, Germany \\
Email: goebel@informatik.uni-hamburg.de}%
\IEEEauthorblockA{%
\IEEEauthorrefmark{2}%
Department of Mathematics and Statistics,
University of Melbourne,
Vic 3010,  Australia \\
Email: \{keeler,taylorpg\}@unimelb.edu.au}%
\IEEEauthorblockA{%
\IEEEauthorrefmark{3}%
Department of Mathematical Sciences,
Stellenbosch University,
7600 Stellenbosch, South Africa \\
Email: aek1@cs.sun.ac.za}%
\thanks{%
The work of Anthony Krzesinski is supported by the Research Foundation
of South Africa (Grant specific unique reference number (UID) 83965) and
Telkom SA Limited.}
\thanks{%
The work of Peter Taylor is supported by the Australian Research Council
Laureate Fellowship FL130100039 and the ARC Centre of Excellence for
Mathematical and Statistical Frontiers (ACEMS).}
}

\begin{document}

\maketitle

\begin{abstract}

In the context of the `selfish-mine' strategy proposed by Eyal and
Sirer, we study the effect of propagation delay on the evolution of the
Bitcoin \blockchain. 
First, we use a simplified Markov model that tracks the contrasting
states of belief about the \blockchain\ of a small pool of miners and
the `rest of the community' to establish that the use of block-hiding
strategies, such as selfish-mine, causes the rate of production of
orphan blocks to increase. 
Then we use a spatial Poisson process model to study values of Eyal and
Sirer's parameter $\gamma$, which denotes the proportion of the honest
community that mine on a previously-secret block released by the pool in
response to the mining of a block by the honest community. 
Finally, we use discrete-event simulation to study the behaviour of a
network of Bitcoin miners, a proportion of which is colluding in using
the selfish-mine strategy, under the assumption that there is a
propagation delay in the communication of information between miners. 

\end{abstract}


\section{Introduction}
\label{sec:intro}

Bitcoin is a peer to peer electronic payment system in which transactions
are performed without the need for a central clearing agency to
authorize transactions.
Bitcoin users conduct transactions by transmitting electronic messages
which identify who is to be debited, who is to be credited, and where
the change (if any) is to be deposited. 

Bitcoin payments use Public Key Encryption. The payers and payees are
identified by the public keys of their Bitcoin wallet identities.  Each
Bitcoin transaction is encrypted and broadcast over the network.
Suppose you receive a transaction from Mary. If you can decrypt Mary's
message using her public key, then you have confirmed that the message
was encrypted using Mary's private key and therefore the message
indisputably came from Mary. But how can you verify that Mary has
sufficient bitcoins to pay you?

The Bitcoin system solves this problem by verifying transactions in a
coded form in a data structure called the {\it blockchain}, which is
maintained by a community of participants, known as {\it miners}. 

It can happen that different miners have different versions of the
\blockchain, something which occurs because of propagation delays, see
Decker and Wattenhofer~\cite{dewa13}. For Bitcoin to be able to
function, it is essential that these inconsistencies are resolved within
a short timescale. We are interested in how the inconsistencies arise
and how they are resolved
(1)~when all participants are acting according to the Bitcoin
protocol, and
(2)~when a pool of participants is using the `selfish-mine' strategy
proposed by Eyal and Sirer~\cite{eysi13}.

\subsection{The \blockchain}
\label{subsec:blockchain}

At the heart of the Bitcoin system is the computational process called
{\it mining}, which involves the solution of a computationally-difficult
cryptographic problem. Bitcoin miners receive copies of all transactions
as they are generated.  They examine the \blockchain\ to investigate the
history of the bitcoins involved in each transaction. If the proposed
transaction has sufficient bitcoin credit, then it is accepted for
incorporation into the block that the miner is currently working on. 

Each transaction is identified with a double SHA-256 hash. Miners gather
transactions together and use their hashes, together with the hash that
is at the current head of the \blockchain, as inputs to the
cryptographic problem. If a miner succeeds in solving the problem, it is
said to have {\it mined} a block that contains records of all the
transactions that were part of the calculation. The miner receives a
reward (currently 25 bitcoins) for accomplishing this, along with a
small transaction fee gathered from each transaction in the block.

The process works as follows. A miner~$M$ computes a block hash~$h$
over a unique ordering
of the hashes of all the transactions that it is intending to
incorporate into its next block $B$.  It also takes as input the block
solution $s_{i-1}$ at the head of its current version of the \blockchain.
Denoting concatenation of strings by the symbol $+$, the cryptographic
problem that $M$ has to solve is: compute a SHA-256 hash
\begin{equation} \label{eq:update}
s_i = \MD(n + h + s_{i-1}),
\end{equation}
such that $s_i$ has at least a specified number~$x$ of leading zeros
where $x \sim 64$.  The string~$n$ is a random ``nonce" value. If $s_i$
does not have at least~$x$ leading zeros, then~$n$ is updated and $s_i$
is recomputed until a solution is found with the required number of
leading zeros.

Once mined, the new block is communicated to the members of the peer
network and, subject to the fine detail of the rules that we shall
discuss in the next section, the new block is added to the \blockchain\
at each peer.  The \blockchain\ thus functions as a public ledger: it
records every Bitcoin payment ever made.

The objective of the designers of the Bitcoin protocol was to keep the
average rate at which blocks are added to the long-term \blockchain\ at
six blocks per hour. To this end, the value of $x$, which reflects the
difficulty of the computational problem inherent in (\ref{eq:update}),
is adjusted after the creation of each set of 2016 new blocks. If the
previous 2016 blocks have been created at an average rate faster than
six  blocks per hour, then the problem is made more difficult, if they
have been created at a slower average rate, then it is made less
difficult. The effect is that the difficulty varies in response to the
total amount of computational power that the community of miners is
applying. 

The test of whether a particular hash has the required number of leading
zeros is a success/failure experiment whose outcome is independent of
previous experiments. Therefore, the number of experiments required for
the first success is geometrically distributed and, given that the
individual success probabilities are very low and the time taken to
perform an experiment is correspondingly very small, the time taken to
achieve a success is very well-modelled by an exponential random
variable.  It is thus reasonable to model block creation instants as a
Poisson process with a constant rate of six per hour. 

The {\it difficulty} of a sequence of blocks is a measure of the amount
of computing effort that was required to generate the sequence. This can
be evaluated in terms of the numbers of leading zeros that were required
when the blocks in the sequence were created. When Bitcoin was started,
miners used PCs to solve the cryptographic puzzle and earn bitcoins. The
difficulty of the puzzle was increased to limit the rate of producing
bitcoins. Miners started using the parallel processing capabilities of
Graphical Processing Units (GPUs) to solve the cryptographic puzzle. The
difficulty of the puzzle was increased again.  Miners started using
General Programmable Field Arrays (GPFAs). The difficulty was increased
yet again. Today miners use Application Specific Integrated Circuit
(ASIC) computers.

Miners communicate by broadcasting newly-discovered blocks via a
peer-to-peer network. Each miner maintains its own version of the
\blockchain\ based upon the communications that it receives and its own
discoveries.  The protocol is designed so that \blockchains\ are locally
updated in such a way that they are identical at each miner or, if they
differ, then the differences will soon be resolved and the
\blockchains\ will become identical. The way that this process works is
explained in the next subsection. 

\subsection{Blockchain rules} 
\label{subsec:rules}

The material discussed here is obtained from~\cite{protocolRules}.  The
{\it main branch} of the \blockchain\ is defined to be the branch with
highest total difficulty. 

\begin{itemize}

\item \textbf{Blocks.}   There are three categories of blocks

\begin{enumerate}

\item Blocks in the main branch: the transactions in these blocks are
considered to be tentatively confirmed.

\item Blocks in side branches off the main branch: these blocks have
tentatively lost the race to be in the main branch.

\item Blocks which do not link into the main branch, because of a
missing predecessor or $n$th-level predecessor.

\end{enumerate}
Blocks in the first two categories form a tree rooted at the very first
block, which is known as the {\it genesis block}, linked by the
reference to the hash of the predecessor block that each block was built
upon. The tree is almost linear with a few short branches off the main
branch.

\item \textbf{Updating the \blockchain.} Consider the situation where a
node learns of a new block. This block could either be mined locally or
have been communicated after being mined at another node. The actions
that the node takes are to:

\begin{enumerate}

\item Reject the new block if a duplicate of the block is present in any
of the three block categories mentioned above.

\item Check if the predecessor block (that is, the block matching the
previous hash) is in the main branch or a side branch. If it is in
neither, query the peer that sent the new block to ask it to send the
predecessor block.

\item If the predecessor block is in the main branch or a side branch,
add the new block to the \blockchain. There are three cases.

\begin{enumerate}

\item The new block extends the main branch: add the new block to the
main branch. If the new block is mined locally, relay the block to the
node's peers.

\item The new block extends a side branch but does not add enough
difficulty to cause it to become the new main branch: add the new block
to the side branch.

\item The new block extends a side branch which becomes the new main
branch: add the new block to the side branch and

\begin{enumerate}

\item find the fork block on the main branch from which this side branch
forks off,

\item redefine the main branch to extend only to this fork block,

\item add each block on the side branch, from the child of the fork
block to the leaf, to the main branch,

\item delete each block in the old main branch, from the child of the
fork block to the leaf,

\item relay the new block to the node's peers.

\end{enumerate}

\end{enumerate}

\item  Run all these steps (including this one) recursively, for each
block for which the new block is its previous block.

\end{enumerate}

\end{itemize}

\subsection{Blockchain dynamics}
\label{subsec:dynamics}

Suppose miner $M_i$ is mining block $B_i$ with hash $h_i$ on its
version $C$ of the \blockchain\ which has $s_{i-1}$ as its previous
hash, and computes a solution $s_i$ to the cryptographic puzzle with
nonce $n_i$. Miner $M_i$ will add $B_i$ to $C$ and broadcast ($B_i, n_i,
h_i, s_i$) to the network. When another miner $M_j$, who is also working
on the \blockchain\ $C$, receives the communication, it will compute 
\[ s^\prime = \MD(n_i + h_i + s_{j-1}).  \] 

\begin{figure}[!t]
\begin{center}
\resizebox{\columnwidth}{!}%
{\includegraphics{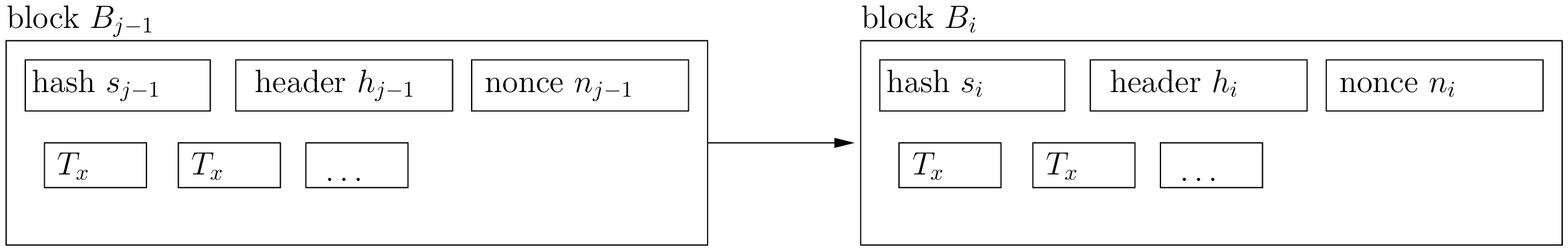}} % blockchain.eps
\end{center}
\caption{Mining a block.}
\label{fig:1}
\end{figure}
With reference to Figure~\ref{fig:1}, if $s^\prime = s_i$ then miner
$M_j$ will add block $B_i$ to its \blockchain\ $C$, abandon the block
$B_j$ that it is working on and commence trying to add a block to the
chain $CB_i$. Any transactions in $B_j$ that are not in $B_i$ will be
incorporated into in this new block. Importantly, miners $M_i$ and $M_j$
now have identical versions of the \blockchain.

The existence of propagation delays can upset the above process, because
blocks can be discovered while communication and validation is in
process. Decker and Wattenhofer~\cite{dewa13} measured the difference
between the time that a node announced the discovery of a new block or a
transaction and the time that it was received by other nodes for a
period of operation in the actual Bitcoin network. They observed that
the median time until a node receives a block was 6.5 seconds, the mean
was 12.6 seconds and the $95$th percentile of the distribution was
around 40 seconds. Moreover, they showed that an exponential
distribution provides a reasonable fit to the propagation delay
distribution. 

Suppose all miners are working on the same version $C$ of the
\blockchain\ and miner $M_i$ mines block $B_i$ at time $t$. It will then
add $B_i$ to the \blockchain\ $C$ and broadcast block $B_i$ to all its
peers. Suppose that this communication reaches miner $M_j$ at time
$t+\delta_j$ and that $M_j$ has mined a block $B_j$ at time $t' \in
[t,t+\delta_j)$.

Miner $M_j$ now knows about two versions $CB_i$ and $C_Bj$ of the
blockchain, which are of the same length.  From the point of view of
Miner $M_j$, the blockchain has split, and we can think of the node as
being in a `race' to see which version of the \blockchain\ survives.

Miner $M_i$ will build on $CB_i$ because this is the version of the
\blockchain\ that it knew about first. However miner, $M_j$ knew about
$CB_j$ first, and will attempt to build on this version of the
\blockchain. Other miners will work on either $CB_i$ or $CB_j$ depending
on which version they heard about first. The `race' situation is
resolved when the next block $B^*$ is mined, say on $CB_i$, and
communicated via the peer network. Then $CB_iB^*$ will be longer than
$CB_j$ and all miners will eventually start building on $CB_iB^*$.
It is then likely that the block $B_j$ will not be part of the longterm
\blockchain\ and it will become an \textit{orphan block}. Any
transactions that are in $B_j$, but not in $B_i$ or $B^*$, will be
incorporated into a future block.'

The above situation can get more complicated if yet more blocks are
mined while communication is taking place, although this would require
the conjunction of two or more low-probability events.

A rough calculation based upon the fact that it takes $600$ seconds on
average for the community to mine a block shows that we should expect
that the probability that a new block is discovered while communication
and validation of a block discovery is taking place is of the order of
$12.6/600 \approx 1/50$, which is small but not negligible. Given that,
on average, 144 blocks are mined each day, we should expect this
circumstance to occur two to three times each day, which accords with
the observed rate of orphan blocks~\cite{orphanblocks}. 

\subsection{Transaction integrity}
\label{subsec:integrity}

In his seminal paper proposing the Bitcoin system~\cite{nakamoto},
Nakamoto dealt with the issue of transaction integrity. He proposed that
a vendor should wait until his/her payment transaction has been included
in a block, and then $z$ further blocks have been added to the
\blockchain, before dispatching the purchased goods. The rule-of-thumb
that has been adopted is to take $z=6$, which roughly corresponds to
waiting for an hour before dispatching the goods.  Assuming that the
community can generate blocks at rate $\lambda_2$, Nakamoto presented a
calculation of the probability $P_A$ that an attacker with enough
computing power to generate blocks at rate $\lambda_1< \lambda_2$ could
rewrite the history of the payment transaction by creating an alternate
version of the \blockchain\ that is longer than the community's version.
Unfortunately, Nakamoto's calculation is incorrect, a fact that was
observed by Rosenfeld in~\cite{rose14}.

Let the random variable $K$ be the number of blocks created by the
attacker in the time that it takes the community to create $z$ blocks.
Then, we can get the correct expression for the probability that the
attack is successful by noting that $z+K$ is the number of Bernoulli
trials required to achieve $z$ successes, with the success probability
of an individual trial given by $p\equiv
\lambda_2/(\lambda_1+\lambda_2)$. It is thus a {\it negative binomial
random variable} with parameters $p$ and $z$. 

Now, using Nakamoto's observation that, conditional on the attacker
having created $K$ blocks when the vendor dispatches the goods, the
probability of the attacker ever being able to build a \blockchain\
longer than the community \blockchain\ is
$\left(\lambda_1/\lambda_2\right)^{z-K}$ if $K<z$, and one otherwise, we
arrive at the expression in equation~(1) of~\cite{rose14},
\begin{equation}
\label{eq:catchup}
P_A = 1-\sum_{k=0}^z {{z+k-1}\choose{z-1}}\left(p^z(1-p)^k-p^k(1-p)^z\right).
\end{equation}

\subsection{Selfish-mine}
\label{subsec:dishonest}

It follows from an analysis similar to that in Section
\ref{subsec:integrity} that, if a group of miners control more than half
of the total computer power, they can collude to rewrite the history of
the transactions. There might, however, be ways for a group to gain an
advantage even if it does not control a majority of the computational
power.

In~\cite{eysi13}, Eyal and Sirer proposed a strategy, called
`selfish-mine', and claimed that, using this strategy, a pool of
colluding `dishonest' miners, with a proportion $\alpha<1/2$ of the
total computational power, can earn a proportion greater than $\alpha$
of the mining revenue. In this sense, a pool of miners collaborating in
using the selfish-mine strategy can earn more than its fair share of the
total revenue.

In brief, selfish-mine works as follows. When a pool miner mines a
block, it informs its colluding pool of miners, but not the whole
community of miners. Effectively, the mining pool creates a {\it secret}
extension of its \blockchain,  which it continues to work on. The honest
miners are unaware of the blocks in the secret extension and continue to
mine and to publish their mined blocks and solutions according to the
standard protocol.

The computational power available to the honest miners is greater than
that available to the mining pool. So, with probability one, the public
branch will eventually become as long as the pool's secret extension.
However it is possible that the secret extension will remain longer than
the public branch in the short term.  The mining pool is giving up the
almost certain revenue that it would receive if it published its
recently-mined block in return for a bet that its secret branch will
become long enough for it to take short-term control of the mining
process. 

Specifically, if the lead happens to become two or more, then the pool
can publish a single block every time that the honest community mines a
block, and publish two blocks when its lead is eventually reduced to
one. In this way the pool works on its version of the \blockchain\ while
allowing the honest community to be engaged in a fruitless search for
blocks that have no chance of being included in the long-term
\blockchain.  

The risk to the pool is that, if it has established a lead of exactly
one by mining a block $B_p$, which it has kept secret, and then it is
informed that the community has mined a block $B_h$, the pool may end up
not getting credit for the block $B_p$. To minimise this risk, the
selfish-mine strategy dictates that the pool should publish the block
$B_p$ immediately it hears about $B_h$. The pool continues working on
$B_p$ itself, and it hopes that at least some of the honest community
will also work on $B_p$, so that the pool will get the credit for $B_p$
if an honest miner manages to extend it. 

When Eyal and Sirer~\cite{eysi13} modelled the selfish-mine strategy,
they included a parameter $\gamma$ to denote the proportion of the
honest community that work on $B_p$ after it has been published
according to the scenario described above. They deduced that the pool
can obtain revenue larger than its relative size provided that
\begin{equation}
\label{eq:alphagamma}
\frac{1-\gamma}{3-2\gamma} < \alpha < \frac{1}{2}.
\end{equation}
Eyal and Sirer's analysis did not, however, take propagation delay into
account. Since the honest community has a head start in propagating
$B_h$ before the dishonest miners have even heard about it and then
there is a further propagation delay before $B_p$ reaches other honest
miners, our first intuition was that $\gamma$ is likely to be very low
in the presence of propagation delays. 

In a survey of subversive mining strategies~\cite{courtoisBahack},
Courtois and Bahack state (provisionally) that the claims made for
efficacy of the selfish-mine strategy~\cite{eysi13}, which is one of the
block discarding attacks studied in~\cite{bahack}, are exaggerated.
However, the conclusions presented in~\cite{courtoisBahack} concerning
the selfish-mine attack are not based on experimental or modelling
analysis.

The purpose of the rest of this paper is to propose some simple models
that explicitly take propagation delay into account, which we can use to
compare the behaviour of the Bitcoin network when all miners are
observing the standard protocol with its behaviour if there is a pool
following the selfish-mine strategy. 

In next section, we shall introduce and analyse a simple continuous-time
Markov chain model that tracks the contrasting states of belief of a
`pool' and the `rest of the community' under the assumption that the
pool and the community are physically-separated so that communication
between the pool and the community takes longer than communication
within the pool and within the community. Effectively, we assume that
there is no communication delay within the pool and within the
community. We conclude that the rate of production of orphan blocks is
likely to be much higher when the pool is keeping its newly-discovered
blocks secret. 

In the following Section \ref{sec:poissonmodel}, we study the value of
Eyal and Sirer's parameter $\gamma$ in a model in which pool miners are
distributed according to Poisson processes in the plane and the
propagation delay between two miners is normally distributed with a mean
that depends on the distance between them. 

Finally, in Section \ref{sec:simulation}, we shall report results from a
simulation of a network of 1,000 miners, of which a fraction form a
dishonest pool, again with propagation delays between all miners that
depend on their spatial separation. Some conclusions and further
observations are given in Section \ref{sec:conclusion}.

\section{A simple Markov chain model}
\label{sec:model}

In this section we shall describe and analyse a simple Markovian model
that takes into account the separate states of belief of a `pool of
Bitcoin miners' and the `rest of the community' about the \blockchain.
We assume that communication within the pool and within the community
always happens faster than communication between the pool and the
community, effectively taking the propagation delay for the former type
of communication to be zero. 

Such a dichotomy between immediate communication within both the pool
and community and delayed communication from pool to community and
vice-versa is unlikely to be realistic. However, the model is useful
because it illustrates the effect that block-withholding strategies have
on the rate of \blockchain\ splits. In the following Sections
\ref{sec:poissonmodel} and \ref{sec:simulation}, we shall analyse models
with more realistic assumptions about communication delay.

If the pool and the rest of the community agree about the \blockchain,
then we denote the state by $(0,0)$. On the other hand, if the pool has
built $k$ blocks onto the last `fork block' where it agreed with the
community, and the community has built $\ell$ blocks beyond the fork
block, then we denote the state by $(k,\ell)$. Given the mechanisms that
are in place to resolve inconsistencies, we would expect that states
$(k,\ell)$ for $k$ and $\ell$ greater than one or two would have a very
low probability of occurrence.

\subsection{The pool mines honestly}
\label{subsec:honestmine}

We assume that the pool discovers new blocks at rate $\lambda_1$, while
the rest of the community does so at rate $\lambda_2$, with
$\lambda_2>\lambda_1$. Without paying attention to node locations,
Decker and Wattenhofer~\cite{dewa13} observed that it is reasonable to
model communication delays with exponential random variables.  Since an
exponential assumption also helps with analytic tractability, we make
such an assumption in this first model. Specifically, we assume that the
time that it takes to communicate a discovery of a block from the pool
to the community and vice-versa is exponentially-distributed with
parameter $\mu \gg \lambda_2$. 

If the system is in a state $(k,\ell)$ with $k\not = \ell$, then it
returns to state $(0,0)$ once communication has occurred, because then
the pool and the community will agree about the new state of the
blockchain. However, if $k=\ell\geq 1$, then the pool and the community
have different, but equal length, versions of the \blockchain\ and will
continue mining on the \blockchain\ as they see it. The system therefore
remains in state $(k,k)$ until a new block is discovered. 

The Markov model has transition rates 
\begin{eqnarray}
q((k,\ell),(k+1,\ell))  & = & \lambda_1, \quad k\geq 0, \ell\geq 0 \label{eq:kk+1}\\  
q((k,\ell),(k,\ell+1))  & = & \lambda_2, \quad k\geq 0, \ell\geq 0 \label{eq:ll+1}\\ 
q((k,\ell),(0,0))       & = & \mu,       \quad k\not=   \ell       \label{eq:kl00}\\
q((k,\ell),(k^\prime,\ell^\prime))
                        & = & 0,         \quad  {\rm otherwise.}   \label{eq:klkls}
\end{eqnarray}
The first two types of transition, reflected in (\ref{eq:kk+1}) and
(\ref{eq:ll+1}), occur when the pool (respectively the community) mine a
block, while the third, in (\ref{eq:kl00}), occurs once communication
has occurred when the chain is in a state $(k,\ell)$ with $k \not =
\ell$. This latter rate is a simplification of what could have been
assumed: if $|k-\ell| \geq 2$, there are multiple communication tasks in
progress, reporting the last $|k-\ell|$ block discoveries in the longest
branch and it is only when the communication reporting the discovery of
the final block on the longest branch arrives that the state of the
system returns to $(0,0)$. For the sake of tractability in this simple
first model, this is the only transition that we have taken into
account. As we observed above, states with $|k-\ell| \geq 2$ have a very
low probability of occurrence and we can expect that this modification
will not have a great effect on the stationary distribution.

The equations for the stationary distribution are
\begin{equation}
\label{eq:00}
\pi(0,0)\left(\lambda_1+\lambda_2\right)
=
\sum_{k=0}^\infty\sum_{\ell=0}^\infty \pi(k,\ell) \mu I (k\not = \ell),
\end{equation}
for $k\not=\ell$,
\begin{eqnarray}
\label{eq:knotequall}
\pi(k,\ell)\left(\lambda_1+\lambda_2+\mu\right)
& = &
\pi(k-1,\ell)\lambda_1 I (k>0) \nonumber \\
& + & \pi(k,\ell-1)\lambda_2 I (\ell>0)
\end{eqnarray}
and, for $k=\ell$,
\begin{eqnarray}
\label{eq:kequall}
\pi((k,\ell))\left(\lambda_1+\lambda_2\right)
& = &
\pi(k-1,\ell)\lambda_1 I (k>0) \nonumber \\
& + & \pi(k,\ell-1)\lambda_2 I (\ell>0).
\end{eqnarray}
To express the solution of these equations, we need to define a function
$n(k,\ell;i)$ which denotes the number of paths that start at the origin
and finish at $(k,\ell)$, take steps on the integer lattice in the
directions $(1,0)$ (north) and $(0,1)$ (east) and which contain exactly
$i$ points $(j,j)$ for $j>0$. 

As an example, we can see that $n(3,2;2)=4$ because there are four paths
\begin{eqnarray*}
& & [(0,0),(1,0),(1,1),(2,1),(2,2),(3,2)], \\
& & [(0,0),(1,0),(1,1),(1,2),(2,2),(3,2)], \\
& & [(0,0),(0,1),(1,1),(2,1),(2,2),(3,2)], \mbox{ and} \\
& & [(0,0),(0,1),(1,1),(1,2),(2,2),(3,2)]
\end{eqnarray*}
that link the origin to $(3,2)$, containing two points of the form
$(j,j)$ for $j>0$.

With $n(k,\ell;i) = 0$ for $i>\min(k,\ell)$, $n(k,0;0)=n(0,\ell;0)=1$
for all $k,\ell>0$, the $n(k,\ell;i)$ for $k\ell \not = 0$ are given by
the recursion
\begin{eqnarray}
\label{eq:nkltrecurs}
\lefteqn{n(k,\ell;i) } \nonumber \\
& = & I(k=\ell)\left[n(k-1,\ell;i-1) + n(k,\ell-1;i-1)\right]\nonumber\\
& + & I(k\not=\ell)\left[n(k-1,\ell;i) + n(k,\ell-1;i)\right].
\end{eqnarray}
For $1\leq i \leq k$, the numbers $T(k,i)=n(k,k;i)$ are known in the
literature. They give the number of {\it Grand Dyck paths} from $(0,0)$
to $(2k,0)$ that meet the $x$-axis $i$ times, which is a simple
transformation of our definition. An expression for these
numbers~\cite[Equation 6.22]{deut99} is
\begin{equation}
\label{eq:dyckpath}
T(k,i)=\frac{i2^i {{2k-i}\choose{k}}}{2k-i}.
\end{equation}
For $k\not = \ell$, the numbers $n(k,\ell;i)$ do not appear in the
Encyclopedia of Integer Sequences~\cite{oeis14}, and we are not aware of
a previous instance where they have been used. However, in a private
communication, Trevor Welsh~\cite{wels14}, produced an expression for
$n(k,\ell;i)$ with $k\not = \ell$. He showed that, for $k>\ell$,
\begin{eqnarray}
n(k,\ell;i)  =  n(\ell,k;i)
             =  \frac{(k-\ell+i)2^i{{k+\ell-i}\choose{k}}}{k+\ell-i},
\label{eq:dyckpath1}
\end{eqnarray}
which generalises (\ref{eq:dyckpath}) in an elegant way.

With the numbers $n(k,\ell;i)$ in hand, we are in a position to write
down the stationary distribution of the Markov chain.

\begin{theorem}
The stationary distribution of the Markov chain defined above has the form
\begin{eqnarray}
\label{eq:stationary}
\lefteqn{\pi(k,\ell) = \pi(0,0) \lambda_1^k\lambda_2^\ell} \nonumber \\
& &
\hspace*{-20pt}
\sum_{i=0}^{\min(k,\ell)}
\frac{(|k-\ell|+i)2^i{{k+\ell-i}\choose{k}}}{(k+\ell-i)(\lambda_1+\lambda_2)^{i}(\lambda_1+\lambda_2+\mu)^{k+\ell-i}},
\end{eqnarray}
where $\pi(0,0)$ is determined by normalisation.

\end{theorem}

\begin{IEEEproof}
The result is established by using (\ref{eq:nkltrecurs}) to verify that
(\ref{eq:stationary}) satisfies (\ref{eq:00}), (\ref{eq:knotequall}) and
(\ref{eq:kequall}).
\end{IEEEproof}

For the case where, $\lambda_1=0.6/hr$,  $\lambda_2=5.4/hr$ (which
corresponds to the pool having $10\%$ of the processing power) and
$\mu=285/hr$, corresponding to Decker and Wattenhofer's~\cite{dewa13}
observed average communication delay of $12.6$ seconds, the values of
$\pi(k,\ell)$ for $k,\ell = 0,\ldots,3$ are given in Table~\ref{tab:1}. 

We see that the pool and the community agree about the \blockchain\
97.5\% of the time, the community has a block that the pool is yet to
hear about  for about 1.8\% of the time, the pool has a block that the
community hasn't heard about for 0.2\% of the time, while the pool and
the community have versions of the \blockchain\ with a single different
final block about 0.4\% of the time. All other possibilities have a
stationary probability less than $10^{-3}$, which supports the intuition
that splits in the \blockchain\ with branches of length greater than one
occur with low probability.

\begin{table}[!t]
\caption{The stationary probabilities $\pi(k,\ell)$ for $k,\ell =
0,\ldots,3$, when the pool mines honestly.}
\label{tab:1}
\centering
\begin{tabular}{|l|cccc|} \hline
  $(k,\ell)$ & 0    & 1  & 2     & 3 \\ \hline
0 & 0.9757 &  0.0181  &  0.0003  &  0.0000\\
1 & 0.0020 &  0.0037  &  0.0001  &  0.0000\\
2 & 0.0000 &  0.0000  &  0.0000  &  0.0000\\
3 & 0.0000 &  0.0000  &  0.0000  &  0.0000\\
\hline
\end{tabular}
\end{table}

Each time that the \blockchain\ is in a state $(1,1)$ and a new block is
mined, approximately one orphan block will be created. This is because
the new state will become $(1,2)$ or $(2,1)$ and, with high-probability,
no other state change will occur before the successful communication
returns the state to $(0,0)$. The block on the shorter branch will then
become an orphan block. With these parameter values, the rate of
creation of orphan blocks is approximately
$\pi(1,1)(\lambda_1+\lambda_2) = 0.022$ per hour, which translates to an
average of about $0.53$ per day. 

Readers will note that this value is much less than the average number
of orphan blocks that are observed each day in the real Bitcoin network,
which lies between two and three.  The discrepancy is explained by the
fact that, in this simple model, we have assumed instantaneous
communication within the pool and within the community. We have not
counted orphan blocks caused by communication delays within the pool and
within the community, which occur in the real network. However, we
believe that the model still has interest because, as we shall see in
Section \ref{subsec:selfishmine}, it can be used to demonstrate that the
rate of production of orphan blocks becomes much higher if the pool is
using a block-hiding strategy such as selfish-mine.

\subsection{The pool uses the selfish-mine strategy}
\label{subsec:selfishmine}

Now we assume that the pool is using the selfish mine strategy described
by Eyal and Sirer in~\cite{eysi13}. As in the model of Section
\ref{subsec:honestmine}, we assume that the pool discovers blocks at
rate $\lambda_1$ and the community discovers blocks at rate $\lambda_2$,
with $\lambda_1< \lambda_2$, independently of the state. 

Under the selfish-mine strategy, the pool does not necessarily publish
blocks immediately it discovers them. Rather, it keeps them secret until
it finds out that the community has discovered a block, and then
publishes one or more of its blocks in response to this news. Most
commonly, this will occur when the pool has a single block $B_p$ that it
has kept secret from the community and then it is notified that the
community has discovered a block $B_h$. The pool's response to this news
is immediately to publish $B_p$, hoping that some of the community will
mine on it. Whether or not this happens, the pool will keep mining on
its own version of the \blockchain. The situation resolves itself when
the next block is discovered, and the state becomes either $(2,1)$ or
$(1,2)$, in which case, with high probability, the state will revert to
$(0,0)$ once communication has taken place. 

Since we have assumed that communication is instantaneous within the
pool and community, but takes time from one to the other, Eyal and
Sirer's parameter $\gamma$, the proportion of the honest community that
mines on the pool's recently-released block when the state is $(1,1)$,
is equal to zero. Thus, when the state is $(1,1)$, a new block will be
created on the pool's leaf at rate $\lambda_1$ and on the community's
leaf at rate $\lambda_2$. 

If the pool has a lead that is greater than or equal to three (a rare
occurrence), it does nothing until it is notified of the discovery of a
block by the community. It then publishes its first block. However,
since the pool and the community will still keep working on the blocks
at the ends of their respective branches, this does not affect the state
of the system, and therefore we put  $q((k,\ell),(0,0))  =  0$ when
$\ell \leq k-2$. 

If the pool has a lead of exactly two and it is notified of the
discovery of a block by the community, the system moves to state $(2,1)$
(or, indeed, the very unlikely states $(3,2)$, $(4,3)$ etc.), and then
the pool will publish all its blocks. Once the communication of the
final block has occurred, the rest of the community will start working
on the longer pool branch, thus returning the state of the system to
$(0,0)$. When it publishes blocks in this situation, the pool is
`cashing-in' on the lead that it has built up, rendering useless the
work that the community has been doing on its branch. This behaviour is
reflected in our Markov model by putting $q((k,k-1),(0,0)) = \mu$ when
$k \geq 2$, where the time taken to communicate a block from the pool to
the community and vice-versa is exponentially-distributed with parameter
$\mu \gg \lambda_2$. 

Finally, we have  $q((k,\ell),(0,0))  =  \mu$ when $k<\ell$, because the
honest miners always publish blocks that they discover, and the pool has
no choice but to build on the community's version of the \blockchain\ if
it is longer. As above and in Section \ref{subsec:honestmine}, we are
taking into account only the communication that reports the discovery of
the final block in the community's chain in assigning this transition
rate.

Our model of the state of the \blockchain\ when the pool is using the
selfish-mine strategy has transition rates
\begin{eqnarray}
q((k,\ell),(k+1,\ell))  & = & \lambda_1, \quad k\geq 0, \ell\geq 0,\label{eq:kk+1s}\\  
q((k,\ell),(k,\ell+1))  & = & \lambda_2, \quad k\geq 0, \ell\geq 0,\label{ll+1s}\\ 
q((k,\ell),(0,0))       & = & \mu,       \quad  k<\ell, \label{kl00s}\\
q((k,k-1),(0,0))        & = & \mu,       \quad k\geq 2, \label{kk-100s}\\
q((k,\ell),(k^\prime,\ell^\prime))
                        & = &  0,        \quad  {\rm otherwise.} \label{klkls}
\end{eqnarray}
The equations for the stationary distribution are
\begin{eqnarray}
\label{eq:00s}
\pi(0,0)\left(\lambda_1+\lambda_2\right)
& = &
\sum_{k=0}^\infty\sum_{\ell=k+1}^\infty \pi(k,\ell)\mu \nonumber \\
& + & \sum_{k=2}^\infty\pi(k,k-1)\mu,
\end{eqnarray}
for $\ell>k$,
\begin{eqnarray}
\label{eq:klesss}
\pi(k,\ell)\left(\lambda_1+\lambda_2+\mu\right)
& = &
\pi(k-1,\ell)\lambda_1 I (k>0) \nonumber \\
& + & \pi(k,\ell-1)\lambda_2 I (\ell>0),
\end{eqnarray}
for $\ell=k$,
\begin{eqnarray}
\label{eq:kequalls}
\pi(k,\ell)\left(\lambda_1+\lambda_2\right)
& = &
\pi(k-1,\ell)\lambda_1 I (k>0) \nonumber \\
& + & \pi(k,\ell-1)\lambda_2 I (\ell>0),
\end{eqnarray}
for $\ell=k-1$,
\begin{eqnarray}
\label{eq:kgreaters}
\pi(k,\ell)\left(\lambda_1+\lambda_2+\mu\right)
& = &
\pi(k-1,\ell)\lambda_1 I (k>0) \nonumber \\
& + & \pi(k,\ell-1)\lambda_2 I (\ell>0)
\end{eqnarray}
and, for $\ell<k$ otherwise,
\begin{eqnarray}
\label{eq:klzeros}
\pi(k,\ell)\left(\lambda_1+\lambda_2\right)
& = &
\pi(k-1,\ell)\lambda_1 \nonumber \\
& + &  \pi(k,\ell-1)\lambda_2 I (\ell>0).
\end{eqnarray}
Like the Markov chain in Section \ref{subsec:honestmine}, this Markov
chain has countably-many states but, unlike the former chain, it does
not appear to be possible to write down a simple closed-form expression
similar to (\ref{eq:stationary}) for its stationary distribution.
However, as we observed in respect of the model of Section
\ref{subsec:honestmine}, the stationary probabilities decay very quickly
to zero as $k$ and $\ell$ increase, and we can get a good approximation
by truncating the state space and augmenting the transition rates in a
physically reasonable way so that the Markov chain remains irreducible.
To get the results that we report below, we truncated the state space so
that only states with $k+\ell \leq 6$ were considered and solved the
resulting linear equations in Matlab. For the same parameters that we
used in the model above, Table~\ref{tab:2} contains the stationary
probabilities for the subset of these states where $k,\ell \leq 3$.  

\begin{table}[!t]
\caption{The stationary probabilities $\pi(k,\ell)$ for $k,\ell =
0,\ldots,3$, when the pool mines selfishly.}
\label{tab:2}
\centering
\begin{tabular}{|l|cccc|} \hline
  $(k,\ell)$ & 0    & 1  & 2     & 3 \\ \hline
0 & 0.8177 &  0.0121  &  0.0002  &  0.0000\\
1 & 0.0818 &  0.0749  &  0.0011  &  0.0000\\
2 & 0.0082 &  0.0002  &  0.0003  &  0.0000\\
3 & 0.0008 &  0.0008  &  0.0000  &  0.0000\\
\hline
\end{tabular}
\end{table}

We see now that the \blockchain\ is in a state where the pool and the
community agree for only 82\% of the time. For about 8\% of the time,
the pool is working on a block that it has kept secret and, for another
7.5\% of the time the pool and the community have separate branches of
length one. As we observed in Section \ref{subsec:honestmine}, each time
that the \blockchain\ is in state $(1,1)$ and a new block is mined, an
orphan block will eventually be created. Also, each time the pool
publishes a block in response to the community finding a block, a
further orphan block is created. The conditions for the latter event
occur with a probability of the order of $10^{-4}$, and we therefore see
that the rate of creation of orphan blocks if the pool is playing the
selfish mine strategy is approximately $\pi(1,1)(\lambda_1+\lambda_2) =
0.4494$ per hour, which is about $10.8$ per day. 

Comparing with the similar calculation in Section
\ref{subsec:honestmine} in which the same parameters $\lambda_1$,
$\lambda_2$ and $\mu$ led to a rate of creation of orphan blocks of
$0.5$ per day, this illustrates that the increased rate of orphan block
creation has the potential to be used as a diagnostic tool as to whether
there is a pool of miners that have adopted the selfish-mine strategy.
Specifically, the community can monitor whether a significant proportion
of the miners is using any type of block-hiding strategy by looking for
increases in the rate of production of orphan blocks. In particular, it
would be possible to detect the presence of a pool of miners
implementing the selfish-mine strategy in this way.

\section{Eyal and Sirer's parameter $\gamma$}
\label{sec:poissonmodel}

In the model of Section \ref{sec:model}, we assumed that the pool and
the community were remote from each other, so that communication within
the pool and within the community could effectively be considered to be
instantaneous, while communication between the pool and community
incurred a delay. This is clearly unrealistic. Indeed, it is likely that
the miners of the pool are distributed throughout the honest community
and that there is delay in communication between any two miners, whether
they are both in the pool or not. 

To illustrate the type of approach that can be taken to model this
situation, we shall make some assumptions about the spatial
relationships and the communication delays between pool miners and
miners in the honest community, and derive some insights about the
behaviour of the \blockchain. While the assumptions would need to be
varied to reflect the characteristics of a mining pool in the actual
Bitcoin network, we believe that the insights hold in general. 

Specifically, we assume that the pool miners are distributed according
to a spatial Poisson point process $\Psi=\{ X_i\}$ with constant
intensity $\nu>0$ over the same region $\mathbb{R}^2$ that contains the
honest miners, so $\Psi$ can be considered a random set of pool miner
locations $\{ X_i\}$. The Poisson process is widely used  for stochastic
models of communication networks, for example, the positioning of
transmitters~\cite{baccelli2009stochastic}.  Although we restrict
ourselves to Euclidean space $\mathbb{R}^2$ for illustration purposes,
Baccelli, Norros and Fabien~\cite{baccelli2011performance} introduced a
general framework using Poisson processes to study peer-to-peer
networks, which was then later used by Baccelli \textit{et al.}~\cite{baccelli2013can} to study the scalability of these networks.
It has been remarked~\cite{baccelli2011performance,baccelli2013can} that
the Poisson process in this model can be defined on other spaces more
suitable for studying networks such as hyperbolic
space~\cite{boguna2010sustaining}, which offers a possible avenue for
further research.

Furthermore, we assume that the communication delay between two Miners
$M_i$ and $M_j$, whether pool or honest, that lie a distance $d_{ij}$
apart is normally distributed with a mean $kd_{ij}$ proportional to this
distance and a constant variance $\sigma^2$, independently of other
transmission delays. 
This assumption does not contradict Decker and Wattenhofer~\cite{dewa13}
who modelled the \textit{unconditional} communication delays with
exponential random variables.

\begin{figure}[!t]
%\centering
\hspace*{-1cm}
\begin{tikzpicture}[thick, scale=0.40]
%auxiliary grid
%defining coordinates
\clip (-12,-5) rectangle (20,6);
\coordinate (A) at (-4,0);
\coordinate (B) at (6.0,-3.5);  
\coordinate (G) at (1,3.96);
%\coordinate (C) at (2,0);
\coordinate (D) at (4,0);
 %\draw[fill opacity=0.8, fill=lightgray!80!]  (0,0) ellipse (8cm and 4cm);
\draw  (0,0) ellipse (8cm and 4cm);
%\draw(0,0) ellipse (10cm and 5.3cm);

%Defining intersection points of the two circles - F is the top point and E is the bottom point
%\coordinate  (F) at (intersection 2 of C1 and C2);
%\coordinate  (E) at (intersection 1 of C1 and C2);
%
%Drawing "black dots"
\draw[fill=black](A)circle(3pt);
\draw[fill=black](B)circle(3pt);
\draw[fill=black](D)circle(3pt);
%the scope enviroment is a local enviroment. We use it so the clip command doens not affect what comes after
\begin{scope}
\draw[clip](A)--(G)--(D)--(A);
\end{scope}
\begin{scope}
\draw[clip](A)--(B)--(D)--(A);
%\draw[clip](A)--(F)--(D)--(E)--(A);

%
%\draw[dash pattern=on 1pt off 1pt](A)circle (10pt);
%\draw[dash pattern=on 1pt off 1pt](D)circle (1.2cm);
%%%%%%
 %\draw[clip] (A) circle (2cm);
 %\draw[fill opacity=0.8, fill=lightgray!80!] (D) circle (10.5cm);
 \end{scope}
%%%%%%%
%labelling
%\draw(0.2,0.1)--(-1.5,2.3) node [above]{ $2\phi_\gamma(u)$};
%\draw(1,1.3)--(2,2.3) node [above]{ $A_\gamma(u)$};
%\draw(10,0)--(11,1.2) node [above]{\small $2\psi_\gamma(u)$};
%\draw(5.5,1) node [above]{\small $u$};
\draw(0,-1)node [above]{$d_{12}$};
\draw(-1.8,2)node [above]{$x_{13}$};
\draw(3,1.8)node [above]{$x_{32}$};
\draw(0.6,-2.7)node [above]{$D_{13}$};
\draw(5.8,-2.1)node [above]{$D_{32}$};

\draw(A)node [below]{$M_1$};
\draw(B)node [right]{$M_3$};
\draw(D)node [right]{$M_2$};

%\draw[latex-latex](A)--(D);
%\draw[latex-latex](A)--(B);
%\draw[latex-latex](B)--(D);

\draw(A)--(D);
%\draw(A)--(B);
%\draw(B)--(D);

%\draw(0,-3)node[thick]{\big|};\draw(11,-3)node[thick]{\big|};
%
%
% \draw[|<->|] (50pt,30ex) -- (5cm,30ex);
%Defining and drawing circle with center in A that passes through point C
\end{tikzpicture}
\caption{$P(D>x)$ is the probability that no pool miner is located in
the ellipse with $x_{13}+x_{32}=x$.}
\label{fig:2}
\end{figure}
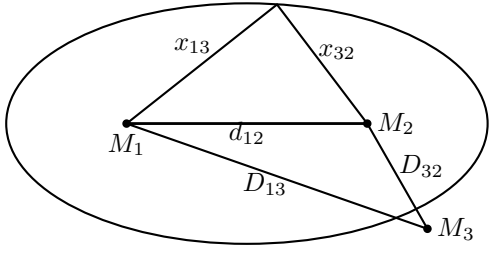

The quantity that we are interested in is Eyal and Sirer's~\cite{eysi13}
proportion $\gamma$ of the honest community that mines on a block
released by the selfish-mine pool in response to the honest community
publishing a block. With reference to Figure~\ref{fig:2}, we are
interested in analysing the communication between two honest Miners
$M_1$ and $M_2$ that lie a distance $d_{12}$ from each other. Miner
$M_3$ is the pool miner for which the length of the path between $M_1$
and $M_2$ via $M_3$ is minimised. Denote the (random) distances between
$M_1$ and $M_3$ and $M_3$ and $M_2$ by $D_{13}$ and $D_{32}$
respectively. 

Consider the situation where the pool has discovered a block $B_p$ that
it has kept secret from the honest community and then honest Miner $M_1$
subsequently discovers and publishes a block $B_h$. The selfish-mine
strategy dictates that Miner $M_3$ should release $B_p$ immediately it
receives $B_h$ from $M_1$. We are interested in the probability that the
other honest Miner $M_2$ will receive $B_p$ before $B_h$ because, with
equal length branches, it will then mine on the branch that it heard
about first.  

Making the further assumption that Miner $M_3$ requires no time to
process the information that a block has arrived from $M_1$ and release
$B_p$ (which could be varied), $\gamma$ is effectively the probability
that communication from $M_1$ to $M_3$ and then $M_3$ to $M_2$ occurs
faster than direct communication from $M_1$ to $M_2$. 

Again with reference to  Figure~\ref{fig:2}, Miner $M_3$ is chosen
so that the distance $D=D_{13}+D_{32}$ is minimal amongst all of the
pool miners. This means that, for any $x < D$, there is no pool miner in
the ellipse whose foci are the locations of honest Miners $M_1$ and
$M_2$ (taken to be at $(-d_{12}/2,0)$ and $(d_{12}/2,0)$ respectively)
and semi-axes 
\begin{equation} \label{eq:ellipse}
a = \frac{x}{2},\qquad
b =\frac{1}{2}(x^2 - d_{12}^2)^{1/2}.
\end{equation}
Hence
\begin{equation}
P(D > x) = e^{-\nu A(x) }, \quad x\geq d_{12},
\end{equation}
where 
\begin{equation} \label{eq:areax}
A(x)= \frac{\pi  x}{4} (x^2 - d_{12}^2)^{1/2}
\end{equation}
is the area of the ellipse (\ref{eq:ellipse}). It follows that
\begin{equation} \label{eq:ddist}
F_D(x) \equiv P(D\leq x) = 1- e^{-\nu A(x) }, \quad x\geq d_{12}.
\end{equation}
Conditional on the random distances $D_{13}$ and $D_{32}$, the
transmission times $T_{13}$ and $T_{32}$ are independent and normally
distributed with means $kD_{13}$ and $kD_{32}$ respectively and common
variance $\sigma^2$, and therefore the difference $\Delta\equiv
T_{13}+T_{32}-T_{12}$ is a normally distributed random variable with
mean $k(D-d_{12})$ and variance $3\sigma^2$. Since the triangle
inequality ensures that the mean of $\Delta$, $k(D-d_{12})$ is
nonnegative, we immediately see that $\widetilde \gamma = P(T < 0)$ is
less than or equal to $0.5$.
Furthermore, 
\begin{equation} \label{eq:condgamma}
P(\Delta<0 |D=x) =  \Phi\left(
\frac{k(d_{12}-x) }{\sqrt{3}\sigma}\right),
\end{equation}
where $\Phi$ is the distribution function of a standard normal random
variable. Integrating with respect to the probability density of $D$
derived from (\ref{eq:ddist}), we see that the probability that the
honest Miner $M_2$ receives $B_p$ before $B_h$ is given by
\begin{equation} \label{eq:gamma}
\widetilde \gamma = \nu
\int_{d_{12}}^{\infty} A'(x) e^{-\nu A(x)} \Phi\left(\frac{k(d_{12}-x)
}{\sqrt{3}\sigma}\right) dx.
\end{equation}
A change of variable $w=A(x)$ results in a numerically tractable Laplace
transform 
\begin{equation} \label{eq:gammalaplace}
\widetilde \gamma = \nu
\int_{0}^{\infty}  e^{-\nu w} \Phi\left( \frac{k(d_{12}-A^{-1}(w))
}{\sqrt{3}\sigma}\right) dw,
\end{equation}
where 
\begin{equation} \label{eq:ainv}
A^{-1}(w)=\frac{1}{\sqrt{2}}([(8w/\pi)^2 +d_{12}^4]^{1/2}
+d_{12}^2)^{1/2}.
\end{equation}

\begin{table}[!t]
\caption{Values of $\widetilde \gamma$ for different values of $d_{12}$
and $\nu$.}
\label{tab:3}
\centering
\begin{tabular}{|l|cccc|} \hline
  $(d_{12},\nu)$ & 0.4    & 0.8  & 1.2    & 1.6 \\ \hline
1  & 0.0341 & 0.0654  & 0.0942  & 0.1207\\
4  & 0.2034 & 0.3144  & 0.3779  & 0.4160\\
8  & 0.3687 & 0.4505  & 0.4758  & 0.4860\\
12 & 0.4430 & 0.4835  & 0.4925  & 0.4958\\
\hline
\end{tabular}
\end{table}

It is clear that $\widetilde \gamma$ depends on $k$ and $\sigma$ only
through the ratio $k/\sigma$. Taking this ratio to be equal to 50,
Table~\ref{tab:3} presents some values of $\widetilde \gamma$
as the distance $d_{12}$ between $M_1$ and $M_2$ and the density $\nu$
of pool miners are varied. We see that, as $d_{12}$ increases, the value
of $\widetilde \gamma$ approaches its theoretical limit of $0.5$. The
rate of convergence is faster if $\nu$ is larger, but the parameter
$\widetilde \gamma$ is more sensitive to the distance $d_{12}$ between
honest Miners $M_1$ and $M_2$ than it is to the intensity of the Poisson
process of pool miner locations. The intuition behind this is that, when
$d_{12}$ is large, there is a high probability that there will be a pool
miner close to the straight line between Miners $M_1$ and $M_2$ even if
the value of $\nu$ is only moderate. 

\begin{figure}[!b]
\begin{center}
\resizebox{\columnwidth}{!}%
{\includegraphics[trim=80 250 0 250]{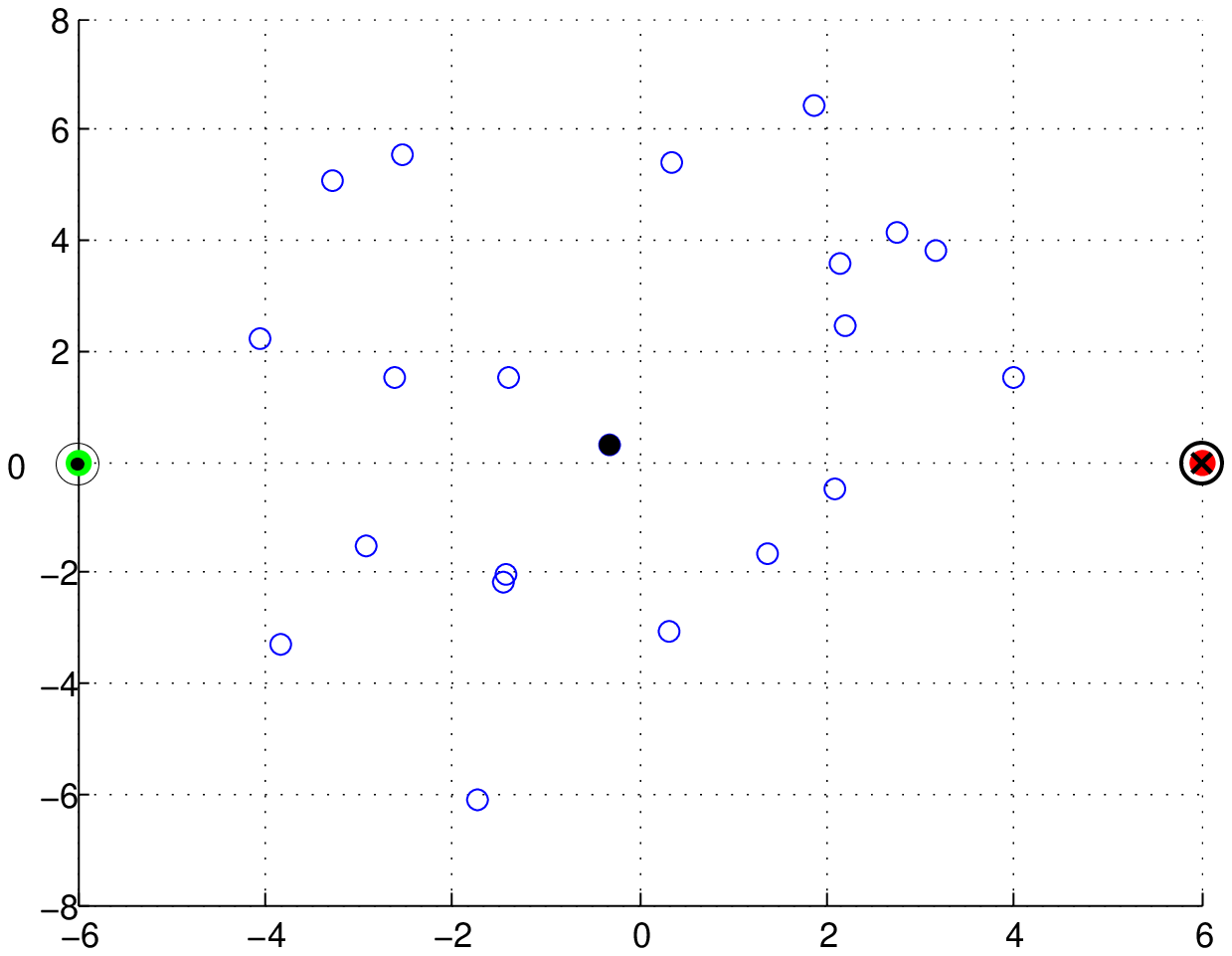}} % PoissonEdited.ps
\end{center}
\caption{An example simulation of the Poisson model with $d_{12}=12$ and $\nu=0.4$.}
\label{fig:3}
\end{figure}

This effect is illustrated in Figure~\ref{fig:3}, which presents
an example where $d_{12}=12$ and $\nu = 0.4$. Honest Miners $M_1$~$\odot$
%(in green)
and $M_2$~$\otimes$
%(in red)
are located at the points $(-6,0)$ and $(6,0)$ respectively.
The round circles $\circ$ are the locations of pool miners, and the
marked pool miner
%in black
$\bullet$ is the pool Miner $M_3$, that minimises the distance
$D_{13}+D_{32}$. Note that, even though the pool miners are not densely
packed, $M_3$ lies very close to the straight line between $M_1$ and
$M_2$.

\begin{figure}[!t]
\begin{center}
\resizebox{\columnwidth}{!}%
{\includegraphics[trim=80 250 0 250]{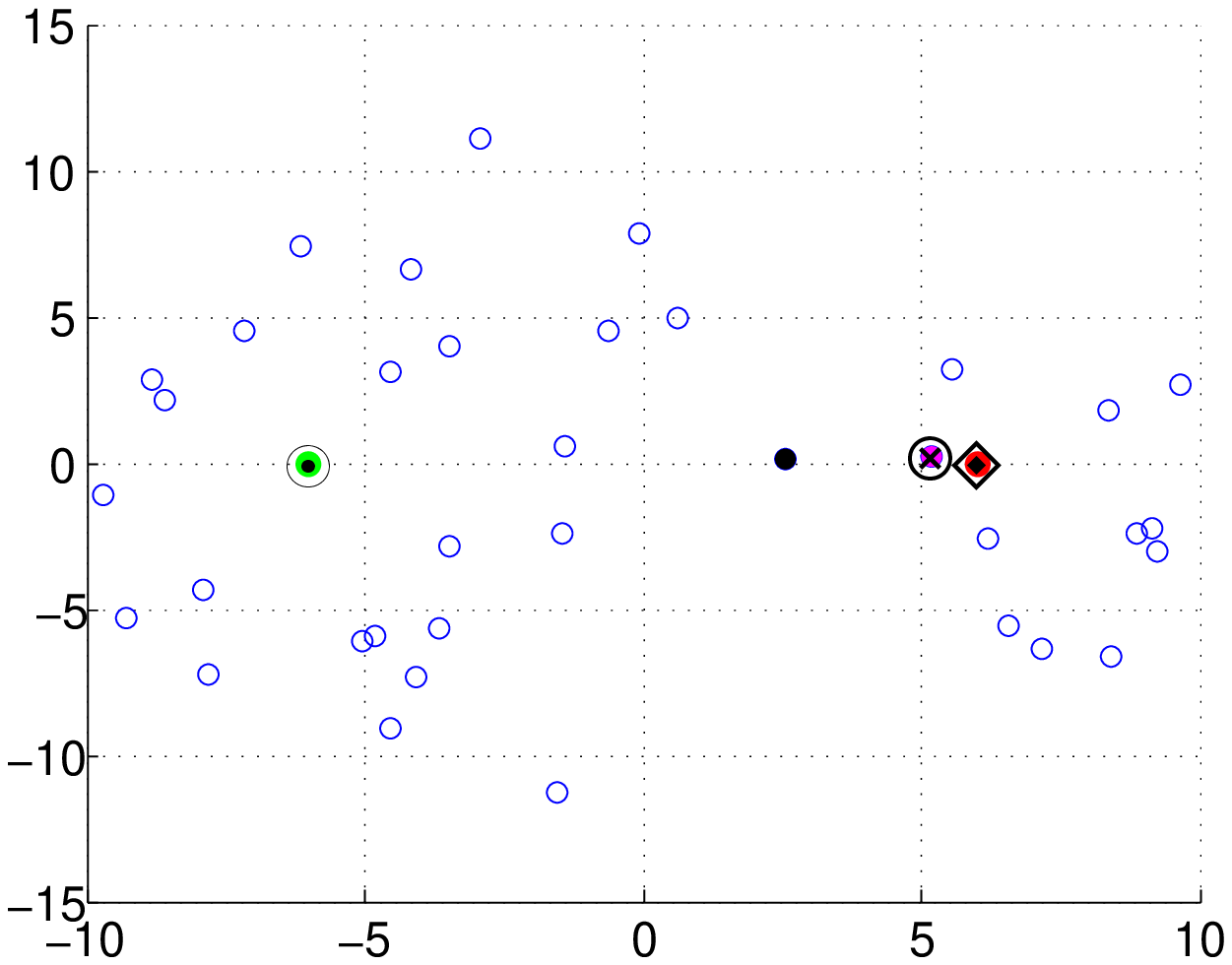}} % Poisson1Edited.ps
\end{center}
\caption{A simulation where the pool node $\otimes$  with the shortest round trip
time is not the pool node $\bullet$ that lies closest to the
straight line between $M_1$ $\odot$ and $M_2$ $\Diamond$; $d_{12}=12$
and $\nu=0.2$.}
\label{fig:4}
\end{figure}

Under the assumptions of the model, the above analysis calculates the
probability that the pool miner $M_3$ closest to the straight line
between honest Miners $M_1$ and $M_2$ succeeds in transmitting $B_p$ to
$M_2$ before $M_2$ directly receives $B_h$. Miner $M_3$ is the pool
miner with the highest probability of succeeding in this transmission.
However, there might be other pool miners that have a round-trip
distance that is not much further than that via $M_3$, and a complete
analysis should take into account the possibility that one of these
miners succeeds when $M_3$ does not. Such a situation is illustrated in
Figure~\ref{fig:4}, where the Miner $M_3$ closest to the straight line
between Miners $M_1$ and $M_2$ is not the miner that had the smallest
value of the round-trip propagation delay.

More precisely, instead of calculating the probability that the
communication time via the pool node $M_3$ that minimises the round-trip
distance is less than the direct transmission time, we should calculate
the probability that the minimum of the communication times via all the
dishonest nodes is less than the direct transmission time. Based upon
our assumptions that the dishonest nodes are distributed as a spatial
Poisson process and that transmission delays are normally-distributed,
the following result helps with this calculation. 

Let $\rho(y)$ denote the distance from honest miner $M_1$ to miner $M_2$
via an intermediate node located at $y\in \mathbb{R}^2$. Then the
distances 
\[
\{D_i\}= \left\{ \rho(X_i)  : X_i\in \Psi \right\}
\]
from Miner $M_1$ to Miner $M_2$ via dishonest users $\{X_i\}= \Psi$ form
a point process on the infinite interval $[d_{12}, \infty)$ and the
following lemma is a consequence of the Mapping Theorem, see, for
example, Kingman~\cite[page 17]{king93}.
\begin{lemma}
The point process $\{D_i\}$ is an inhomogeneous Poisson point
process with intensity (or mean) measure given by
\begin{equation}
\label{eq:LambdaD}
\Lambda_D(x) := \Lambda_D\left(  [  d_{12} , x ]  \right)
= \nu A(x) , \quad x\geq d_{12}
\end{equation}
where $A(x)$ is given by (\ref{eq:areax}).
\end{lemma}

We can make further use of the Mapping Theorem to obtain a lemma about
the Poisson nature of the round trip times.

\begin{lemma} \label{lem:2}
The point process $\{T_i\}$ is an inhomogeneous Poisson point
process on $(-\infty,\infty)$ with intensity measure given by
\begin{equation}
\label{eq:LambdaT}
\Lambda_T(y) := \Lambda_T\left( (-\infty , y ]  \right)
= \nu\int_{d_{12}}^{\infty} A'(x) \Phi\left(\frac{y-kx }{\sqrt 2\sigma}\right) dx,
\end{equation}
where $A(x)$ is given by (\ref{eq:areax}).
\end{lemma}

\begin{IEEEproof}
We can write $T_i = kD_i + E_i$ where the sequence $\{E_i\}$ consists of
i.i.d. $N(0,2\sigma^2)$ random variables, independent of the sequence
$\{D_i\}$, where, in the theory of marked point process, each $E_i$  is
referred to as a random mark. By the Marking Theorem~\cite[page
55]{king93}, the two-dimensional process $(D_i,E_i)$ is also a Poisson
process, with intensity measure on rectangles of the form $(a,b]\times
(-\infty,y]$ given by
\[
\Lambda_{D,E}(a,b,y) = \nu \Phi\left(\frac{y}{\sqrt 2\sigma}\right)  \int_a^b A'(x) dx,
\]
and the Poisson nature of the process $\{T_i\}$ follows again from the
Mapping Theorem~\cite[page 17]{king93}. To get the expression
(\ref{eq:LambdaT}) for the intensity measure of $\{T_i\}$, we condition
on the possible value of $D_i$ that leads to a given value of $T_i$.
\end{IEEEproof}

The probability $\gamma$ that the pool block released by $M_3$ will
reach $M_2$ before the block published by $M_1$ is the probability that
there exists a point of the Poisson process  $\{T_i\}$ less than the
direct transmission time $T_{12}$. This latter time is normally
distributed with mean $kd_{12}$ and variance $\sigma^2$. If such a point
exists, then there will be at least one pool node where the round-trip
time is shorter than the direct time. 

Conditional on $T_{12}=t_{12}$, we can use Lemma \ref{lem:2} to write
the probability of the above event as
\begin{equation}
\label{eq:delayprobt12}
P(\min T_{i}\leq T_{12}|T_{12}=t_{12})= 1- \exp(-\Lambda_T(t_{12})),
\end{equation}
where $\Lambda_T$ is given by $(\ref{eq:LambdaT})$. This expression can
also be derived by considering $\min T_i$ as extremal shot-noise, see,
for example, Baccelli and B\l aszczyszyn~\cite[Proposition
2.13]{baccelli2009stochastic}.

Now, integrating with respect to the density of $T_{12}$, the
unconditional probability that there is a point of the round trip
process which is less than the direct transmission time is
{\small
\begin{eqnarray} \label{eq:gammamin}
\gamma
& = &     \frac{1}{\sqrt{2\pi}\sigma}\int_{-\infty}^{\infty}
\left(1- \exp(-\Lambda_T(u))\right)  \exp\left(\frac{-(u-kd_{12})^2}{2\sigma^2}\right) du \nonumber\\
& = &  1- \frac{1}{\sqrt{2\pi}\sigma}\int_{-\infty}^{\infty}
\exp\left(\frac{-(u-kd_{12})^2}{2\sigma^2} - \Lambda_T(u)\right) du.
\end{eqnarray}
}
For the same values of $d_{12}$ and $\nu$ that were used in
Table~\ref{tab:3}, again with $k/\sigma =  50$, Table \ref{tab:4} gives
the values of $\gamma$ calculated via~(\ref{eq:gammamin}).

\begin{table}[!t]
\caption{Values of $\gamma$ for different values of $d_{12}$ and $\nu$.}
\label{tab:4}
\centering
\begin{tabular}{|l|cccc|} \hline
  $(d_{12},\nu)$ & 0.4    & 0.8  & 1.2    & 1.6 \\ \hline
 1 & 0.0347 & 0.0678  & 0.0992  & 0.1292\\
 4 & 0.2298 & 0.3914  & 0.5081  & 0.5946\\
 8 & 0.4891 & 0.6937  & 0.7955  & 0.8530\\
12 & 0.6695 & 0.8372  & 0.9018  & 0.9336\\
\hline
\end{tabular}
\end{table}

We notice first that the values of $\gamma$ in Table~\ref{tab:4} are
all higher than than the values of $\widetilde \gamma$ depicted in
Table~\ref{tab:3}, reflecting the fact that pool nodes other
than the pool node that is closest to the straight line between $M_1$
and $M_2$ might lie on the path that minimises the round-trip delay.
Furthermore, we see that the values of $\gamma$ are more sensitive to
the density $\nu$ of the pool nodes than the values of $\widetilde
\gamma$. This makes sense because, when the density of pool nodes is
high, there are likely to be more pool nodes, other than the one that
minimises the round-trip distance between $M_1$ and $M_2$, that have
short round-trip times.  Finally, we note that when distance $d_{12}$
between nodes $M_1$ and $M_2$ is high, and the density $\nu$ of pool
nodes is also high, the probability $\gamma$ can be arbitrarily high,
for example exceeding $0.9$ when $d_{12}=12$ and $\nu=1.6$, even though
  the probability $\widetilde \gamma$ cannot be greater than $0.5$. 

The overall lesson from the analysis in this section is that, with
randomly-varying communication delays, it is advantageous for the pool
to maximise the number of nodes that release a secret block in response
to a block being mined by the honest community. This maximises the
probability of at least one of them succeeding in transmitting its
released block to the other honest nodes before they receive the direct
communication from $M_1$.

In fact, a similar observation can also be applied to the honest
community itself. Rather than relying on the direct communication
between $M_1$ and $M_2$ to occur faster than round-trip communication
via the pool nodes, the honest community could also employ intermediate
nodes as relays and there would be a good chance that faster
communication would be achieved via one of these. Analysing such a
situation using the techniques of this section is an interesting
question for future research.  

\section{\Blockchain\ simulation experiments}
\label{sec:simulation}

We developed two \blockchain\ simulators, one in C++ and one in Java,
the latter based on the DESMO-J simulation framework~\cite{pakr05}. We
used the former to simulate a network of 1,000 nodes. The simulation
worked as follows.
\begin{itemize}

\item The positions of the nodes were selected uniformly at random on
the set $[0,1000]\times[0,1000]$. 

\item Blocks were mined at randomly-selected nodes at the instants of a
Poisson process. On average, one block was mined every 10 minutes.

\item Each node maintained a local copy of the \blockchain.

\item The communication delay between two nodes was a random variable
sampled from a normal distribution whose mean was proportional to the
Euclidean distance between the two nodes and whose coefficient of
variation $CV$ was kept constant. Note that this differed from the delay
model described in Section~\ref{sec:poissonmodel}, where we assumed that
the normally distributed communication delay had a constant variance
$\sigma^2$. In the model discussed in this section, the variance
increases with the distance between the nodes.

\item A total of 10,000 blocks were mined. This represents 70 days of
mining.

\item Each simulation experiment was replicated 12 times and 95\%
confidence intervals for all performance measures that we shall discuss
below were computed.

\item The simulation results are generally presented below in the form
of plots.  The plotted points are sample means. Confidence interval half
widths are shown if they are distinguishable, otherwise they are
omitted. The plotted points are connected by continuous curves
constructed from segments of cubic polynomials whose coefficients are
found by weighting the data points.

\end{itemize}

\begin{figure}[!t]
\centering
\resizebox{\columnwidth}{!}{\includegraphics{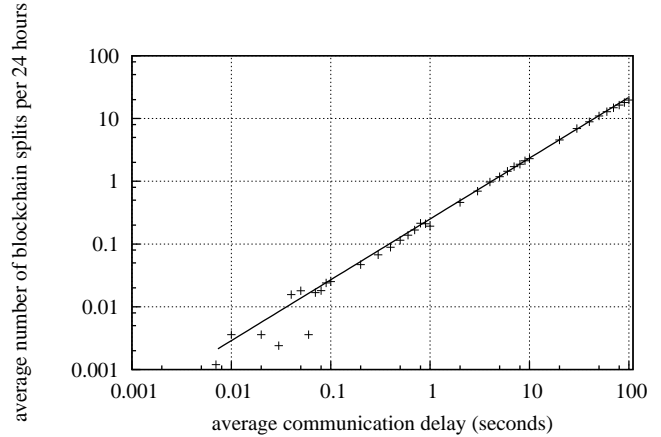}} % fig1a.ps
\caption{The average number of \blockchain\ splits per 24 hours.}
\label{fig:5}
\end{figure}

\subsection{Honest mining}

Figure~\ref{fig:5} shows the average number $b(t)$  of \blockchain\
splits per 24~hours as a function of the communication delay~$t$,
averaged over all the nodes in the network. The delay was varied from
1~msec to 100~seconds. Both axes are logarithmic. Fitting a straight
line to the log-log plot yields $b(t) = 0.2508 t ^{0.9695}$ so that the
average split rate was almost linearly proportional to the average
communication delay.

The simulation experiments showed that when the expected communication
delay was 10~seconds, on average 2.34 splits were observed per 24~hours.
This is roughly in agreement with the observations made by Decker and
Wattenhofer~\cite{dewa13} that an average communication delay of 12.6
seconds results in an average split rate of 2.4 per 24~hours in the
actual Bitcoin network.

\begin{figure}[!b]
\psfrag{average dwell time (seconds)} {\LARGE average dwell time (seconds)}
\psfrag{average communication delay (seconds)} {\LARGE average communication delay (seconds)}
\centering
\resizebox{\columnwidth}{!}{\includegraphics{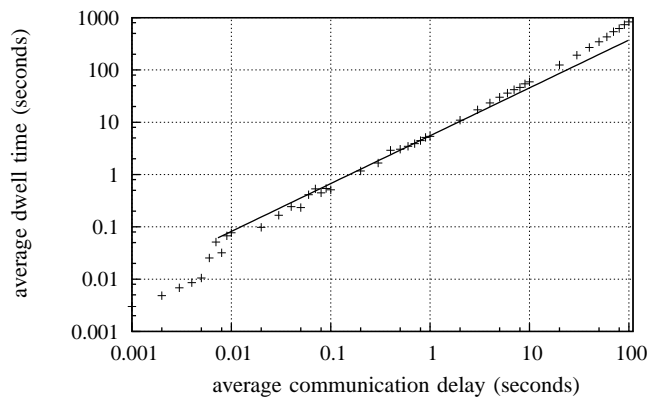}} % fig1c.ps
\caption{The average dwell time.}
\label{fig:6}
\end{figure}

Suppose there is a (hypothetical) mechanism that is invoked when a block
is attached to a \blockchain. The mechanism can simultaneously inspect
the \blockchains\ at all the nodes and report if each \blockchain\ has a
single leaf and if all the \blockchains\ are identical: if this
condition occurs then the \blockchains\ are said to be {\it
synchronised}.

Consider an instant of time~$t_0$ when the mechanism reports that the
\blockchains\ are synchronised.  Let $t > t_0$ denote the first time
instant after~$t_0$ when the mechanism reports that the \blockchains\
are not synchronised.  Let $t' > t$ denote the first time instant
after~$t$ when the mechanism reports that the \blockchains\ are again
synchronised.  We shall refer to the interval $t'-t$ as the
\textit{dwell time}.

Figure~\ref{fig:6} plots the average dwell time as a function of the
average communication delay.  Again, the axes are logarithmic.  The
figure shows that the dwell time was also almost linearly proportional
to the average communication delay.

As the average communication delay increased, the number of splits
increased and the time until the splits were resolved and the
\blockchains\ were synchronised increased.  The average dwell time
exceeded 10~minutes (the average time between mining events) when the
average communication delay was of the order of 100~seconds.

\subsection{Dishonest mining} \label{sec:dishonest}

In remainder of this section we shall report the application of our
simulator to the situation where a pool of miners used Eyal and Sirer's
selfish-mine approach~\cite{eysi13}. The details of our implementation
of the selfish-mine algorithm are given in the Appendix. 

\ifthenelse{\ieeeexplore=1}
{
We have included a supplementary downloadable file available at
\url{http://ieeeexplore.ieee.org} which provides an animation of the
Bitcoin \blockchain\ at a dishonest node in a 1,000-node network with
500 dishonest miners. The animation illustrates the formation of the
secret extension, \blockchain\ splits, race conditions and their
resolution, and the selective publication of secret blocks in response
to the announcement of public blocks.  The file is 218KB in size.
}{}

As in Section \ref{subsec:dishonest}, we use $\alpha$ to denote the
fraction of the total computing capacity of the network that is
controlled by the dishonest pool, and $\gamma$ to denote the probability
that an honest miner will mine on the block $B_p$, rather than $B_h$.

When the communication delays are zero, according to Eyal and Sirer's
expression (\ref{eq:alphagamma}), the minimum proportion of computing
power required for profitable selfish mining ranges from $\alpha > 0$
(if $\gamma=1$) to $\alpha > 1/3$ (if $\gamma=0$).

We simulated the communication delays between miners in the network as
independent normal random variables whose mean was proportional to
distance between the miners and whose coefficient of variation $CV$ was
kept constant. If $CV = 0$ then, by the triangle inequality, we would
have expected $B_h$ to reach $M_2$ before $B_p$ reaches $M_2$, unless
the three nodes $M_1$, $M_2$ and $M_3$ are collinear, which is an event
of probability zero. This expectation was confirmed in the simulation.

\begin{figure}[!t]

\centering
\psfrag{alpha}{\Huge$\alpha$}
\psfrag{gamma}{\rotatebox{-90}{\Huge$\gamma$}}
\resizebox{\columnwidth}{!}{\includegraphics{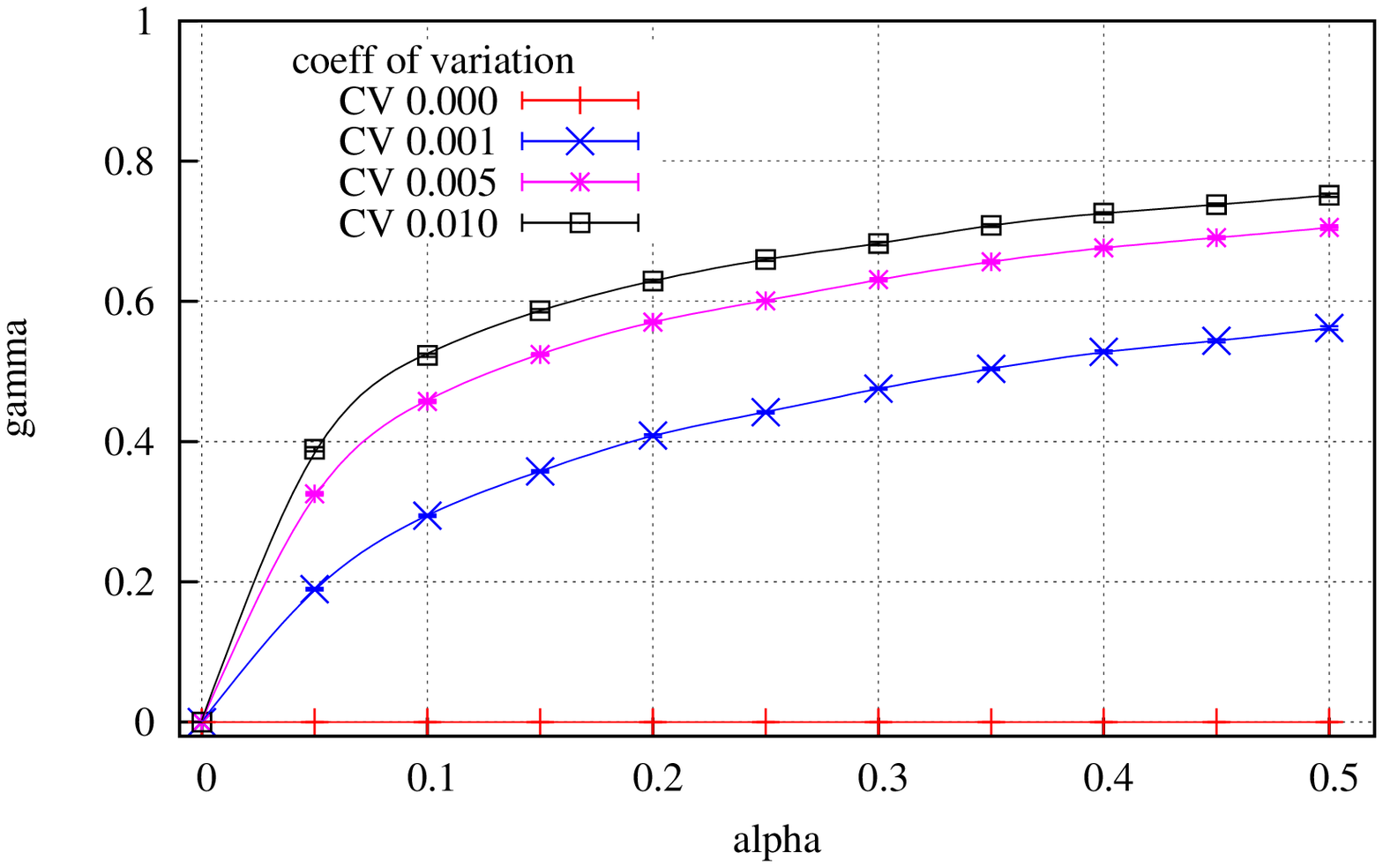}} % fig5a.ps
\caption{The ratio $\gamma$.}
\label{fig:7}
\end{figure}

%~/f/BITCOIN/VERSION6/fig5.gnu
%set loadpath "test_gamma_1000_a40_c10000_C500_e1_g5_t2_v1_V"
\begin{figure}[!t]
\psfrag{alpha}{\Huge$\alpha$}
\psfrag{GAMMA}{\rotatebox{-90}{\Huge$\Gamma$}}
\centering
\resizebox{\columnwidth}{!}{\includegraphics{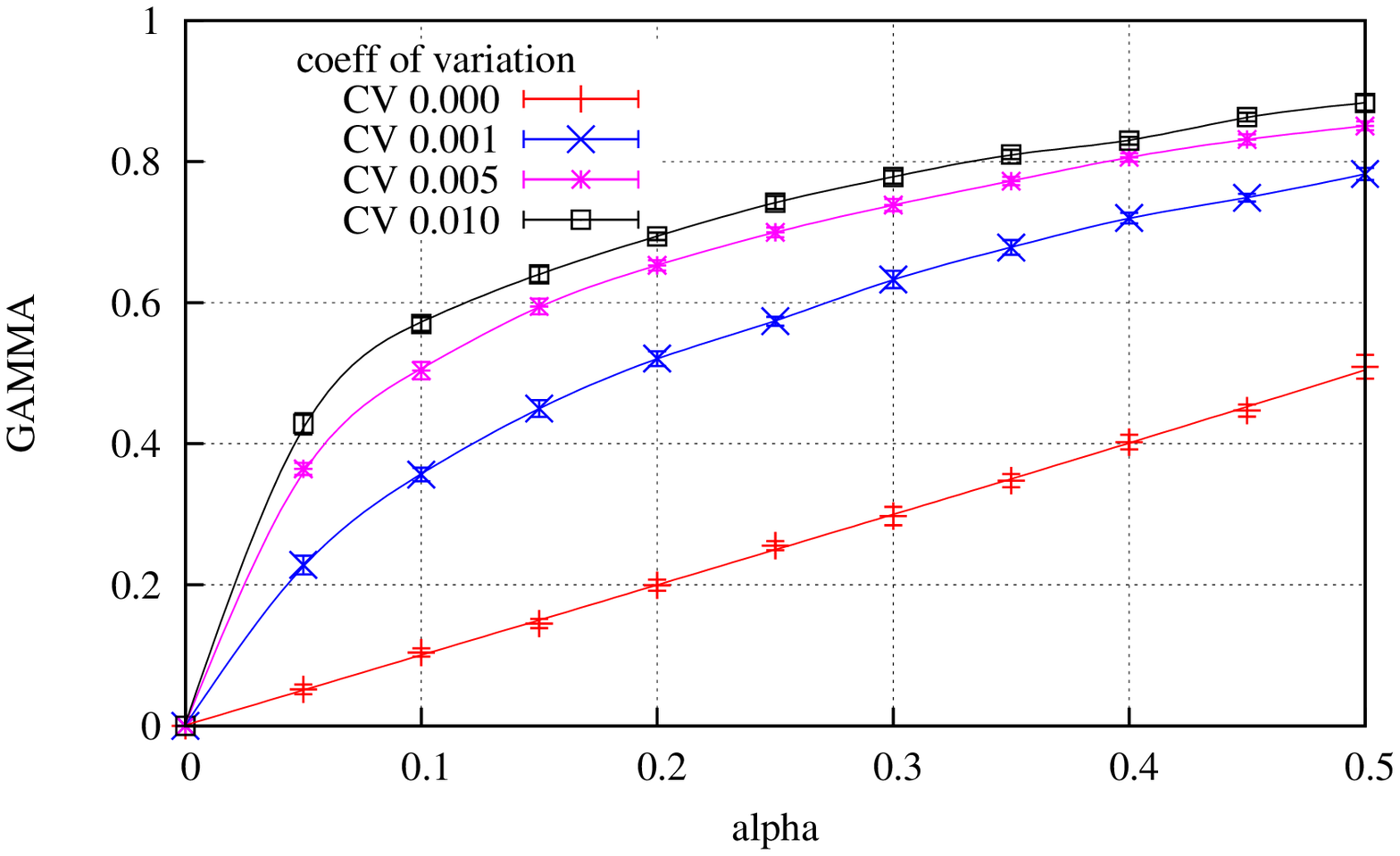}} % fig5b.ps
\caption{The ratio $\Gamma$.}
\label{fig:8}
\end{figure}

However, if $CV > 0$ then $B_p$ can arrive at $M_2$ before $B_h$, and so
$\gamma$ will be positive. We used our simulator to investigate the
value of $\gamma$ as the number of pool miners was varied from from 0 to
500, and thus the proportion $\alpha$ of pool computing power was varied
from $0$ to $0.5$.   Figure~\ref{fig:7} presents the observed proportion
$\widehat \gamma$ as a function of $\alpha$ for several values of the
coefficient of variation~$CV$.  The figure confirms our expectation
that, when $CV > 0$ and there are dishonest miners present, then
$\widehat \gamma$ ought to be positive.

Furthermore, the value of $\widehat\gamma$ increased quickly as a function
of $\alpha$ even when $CV$ was taken to be quite small. This reinforces
the insight that we gained in Section \ref{sec:poissonmodel} that,
because there were many possibilities for the intermediate pool node,
the probability of a communication path via one of them beating the
direct communication was unexpectedly high.

The fact that an honest miner~$M_i$ is mining on a block $B_p$ revealed
by the dishonest pool does not guarantee that the next block to be
attached to the \blockchain\ $C_i$  at node~$M_i$ will be linked to
block $B_p$.
If $C_i$ has two leaves $B_p$ and $B_h$ and a block $B_{new}$ arrives
from another node, then $B_{new}$ can attach to $B_p$ or to $B_h$.

Let $\Gamma$ denote the probability that the next block attached to the
\blockchain\ at an honest node was linked to block $B_p$.
Figure~\ref{fig:8} shows that the sample means of $\Gamma$ indicated by
the points $(+,\times,\ast,\Box)$ corresponded closely with the
theoretical value $\Gamma = \alpha + (1 - \alpha) \gamma$ given by the
continuous curves.

\subsection{The relative pool revenue}

Let $N_h$ denote the total number of blocks mined by the honest miners
that were included in the \blockchain\ at the end of the experiment. The
revenue earned from these blocks has been credited to the honest miners.
Let $N_p$ denote the total number of blocks mined by the pool that were
finally included in the \blockchain. Define the relative pool revenue
\mbox{$R = N_p / (N_h + N_p)$.}

\begin{figure}[!t]
\psfrag{alpha}{\Huge$\alpha$}
\psfrag{number of nodes}{\LARGE number of nodes}
\centering
\resizebox{1.05\columnwidth}{!}{\includegraphics{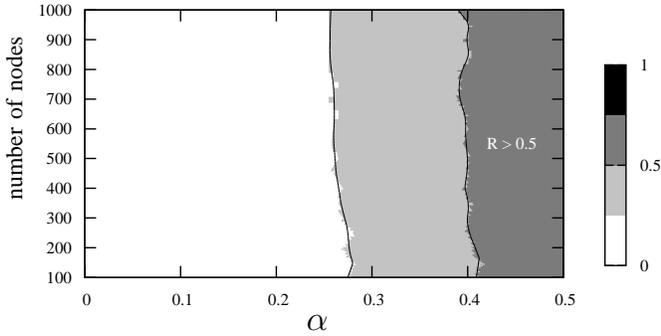}} % figXb.eps
\caption{The relative pool revenue~$R$ in networks of increasing size.}
\label{fig:9}
\end{figure}

\begin{figure}[!b]
\psfrag{alpha}{\Huge$\alpha$}
\psfrag{average number of splits per 24 hours}
{\LARGE average number of splits per 24 hours}
\centering
\resizebox{0.9\columnwidth}{!}{
{\includegraphics{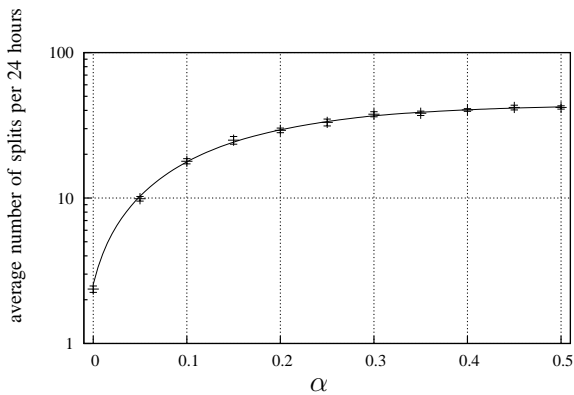}}} % fig7a.ps
\caption{The average number of \blockchain\ splits per 24 hours.}
\label{fig:10}
\end{figure}

Figure~\ref{fig:9} presents a map of the relative pool revenue~$R$ as a
function of the total number of miners (varied from 100 to 1,000) and
the pool size as a fraction $\alpha$ of the total number, with the
average communication delay fixed at 10~seconds.  The figure
demonstrates that the relative pool revenue was roughly constant with
respect to the number of nodes, and increased with increasing values of
$\alpha$. Significantly, $R$ became greater than $0.5$ when $\alpha$
reached $0.4$, which indicates that the pool was earning more than its
fair share of revenue in this region.

\subsection{Detecting the presence of dishonest miners}

In this section and the following sections, we follow the theme of
Section \ref{sec:model}, and discuss how the honest miners can detect
the presence of a pool of miners implementing the selfish-mine strategy.
Consider a network of 1,000 miners, with an average communication delay
of 10~seconds and a coefficient of variation $CV=0.001$.

\begin{figure}[!t]
\centering
\psfrag{alpha}{\Huge$\alpha$}
\psfrag{revenue per miner (bitcoins/hour)}
{\LARGE revenue per miner (bitcoins/hour)}
\resizebox{0.9\columnwidth}{!}{\includegraphics{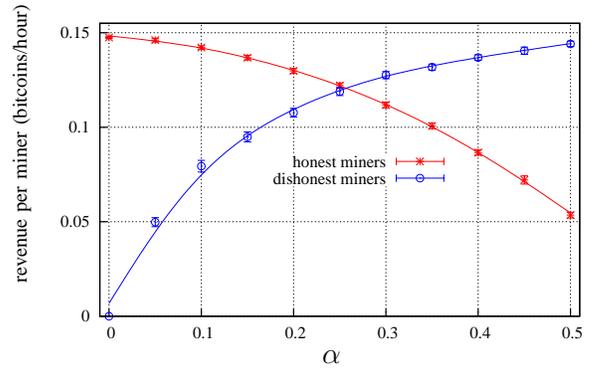}} % fig7c.ps
\caption{The average confirmed revenue earned per hour per miner.}
\label{fig:11}
\end{figure}

Figure~\ref{fig:10} presents the average number of \blockchain\ splits
per 24~hours as a function of the relative size $\alpha$ of the
dishonest pool.  As the size of the dishonest pool increased, the
average number of splits per unit time increased by an order of
magnitude.  Thus the simulation has confirmed the conclusion of the
model that we investigated in Section \ref{sec:model} that an increase
in the split rate can provide a means for the honest miners to detect
the presence of the dishonest miners.

In a network of 1,000 nodes, assuming that all miners have the same
computational power, each miner expects to earn an average of $25 \times
6 / 1000 = 0.15$ bitcoins per hour.  Figure~\ref{fig:11} shows that as
the number of dishonest miners increased, the honest miners earned less
than the expected average of 0.15 bitcoins per hour. This may also
afford a means for the honest miners to detect the presence of a pool of
miners implementing the selfish-mine strategy.

\begin{figure}[!b]
\psfrag{alpha}{\Huge$\alpha$}
\psfrag{relative pool revenue}{\LARGE relative pool revenue}
\centering
\resizebox{\columnwidth}{!}{\includegraphics{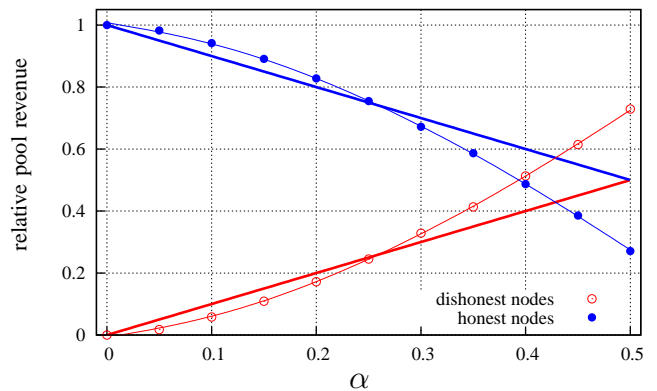}} % fig7d.ps
\caption{The relative pool revenue~$R$ and $\widehat{R}$.}
\label{fig:12}
\end{figure}

\subsection{Dishonest mining is not profitable}
\label{sec:notProfitable}

Figure~\ref{fig:12} presents the relative pool revenue~$R$ as a function
of the relative size $\alpha$ of the dishonest pool. The figure shows
that for $\alpha > 0.25$ dishonest mining outperformed honest mining.
However, this does not imply that the pool incorporated more blocks into
the main branch than it would have if the dishonest miners had followed
the bitcoin rules.

\begin{figure}[!t]
\centering
\psfrag{alpha}{\Huge$\alpha$}
\psfrag{blocks mined per hour}{\LARGE blocks mined per hour}
\resizebox{\columnwidth}{!}{\includegraphics{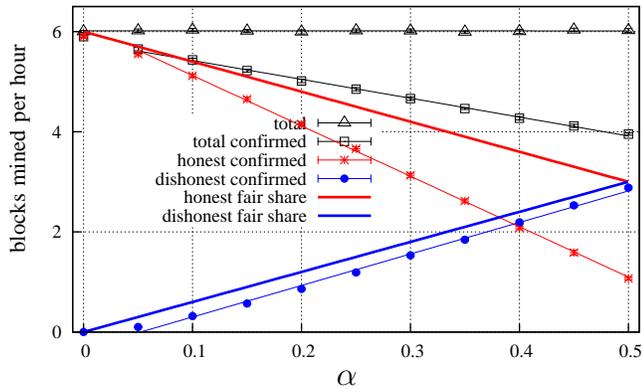}} % fig7b.ps
\caption{The average block mining rate.}
\label{fig:13}
\end{figure}

Figure~\ref{fig:13} illustrates this by exhibiting the performance of
both the dishonest pool and the honest miners in terms of the numbers of
blocks they mined that end up in the main branch. It presents the
average number of blocks mined per hour by the pool, by the honest
miners and in total, that were included in the long-term \blockchain\ as
a function of the relative size $\alpha$ of the dishonest pool.  The
average block mining rate was held constant at 6~blocks per hour.

The figure demonstrates that, when there is a pool implementing
selfish-mine, both the pool and the honest miners were worse off than
they would have been if no dishonest mining was present. The total
number of blocks that the pool and honest nodes incorporated into the
main branch when dishonest mining was present was always less than the
number that would have been incorporated if dishonest mining were not
present.

We caution that the above observation is made under the assumption that
the difficulty of the cryptographic problem described in Section
\ref{subsec:blockchain} was held constant. In the real Bitcoin
\blockchain, the network would respond to an overall decrease in the
rate of blocks being successfully mined by reducing the difficulty of
the cryptographic problem. This decreased value of the difficulty may
itself afford a means for the honest miners to detect the presence of
the dishonest miners.

\subsection{Adoption threshold}

Figure~\ref{fig:12} shows that in the range $0 < \alpha \leq 0.25$ there
is no incentive for a solo miner to adopt the selfish-mine strategy,
since by doing so a miner will be become part of a pool that has a lower
relative pool revenue than it would have if all the members of the pool
were honest.
Moreover, in the range $0 < \alpha \leq 0.25$, solo honest miners
benefit (in terms of their relative pool revenue $\widehat{R} = 1 - R$)
from the activities of the dishonest pool.
A larger participant may already possess more than 25\% of the network
mining capacity and may be able to attract miners with a promise of an
enhanced pool revenue. However, as shown in
Section~\ref{sec:notProfitable}, these miners will earn less than they
would have earned had they remained honest, and if they perceive this
they may withdraw from the dishonest pool.

\section{Conclusions} \label{sec:conclusion}

In this paper we have studied the dynamics of the Bitcoin \blockchain\
when propagation delays are taken into account, with specific reference
to how the \blockchain\ behaves when there is a pool of miners using the
selfish-mine strategy proposed by Eyal and Sirer~\cite{eysi13}. Our
approach has been to construct simple models that provide insight into
system behaviour, without attempting to reflect the detailed structure
of the Bitcoin network.

In Section \ref{sec:model} we used a simple Markov chain model to
demonstrate that it is possible for the whole mining community to detect
block-hiding behaviour, such as that used in selfish-mine, by monitoring
the rate of production of orphan blocks. 

In Section \ref{sec:poissonmodel}, our attention turned to Eyal and
Sirer's parameter $\gamma$, which is the proportion of the honest
community that mine on a previously-secret block released by the
selfish-mine pool in response to the honest community mining a block.
When there is no variability in the propagation delay, it follows from
the triangle equality (at least within the Poisson network model that we
assumed) that the value of $\gamma$ is zero. However, the value of
$\gamma$ can increase surprisingly quickly with increasing variability
of propagation delay. A key observation is that if all pool nodes
release the secret block as soon as they are notified of the discovery
of the public block, the chances of one of them beating the direct
communication can be very high. We did not study the counter-balancing
effect that would occur if all honest miners relayed the honest miner's
block in the same way. A study of this would be an interesting topic for
future research.

Finally, in Section \ref{sec:simulation}, we used simulation to verify
the observations that we made in Sections \ref{sec:model} and
\ref{sec:poissonmodel}, under slightly different assumptions. We also
were able to study the long-term rate of block production, and hence
revenue generation under both honest mining and selfish-mine strategies
and make some observations about when selfish-mine is profitable. An
important observation is that, in the absence of a relaxation of the
difficulty of the mining cryptographic problem, the long-term rate of
block production will decrease if a pool of miners is implementing the
selfish-mine strategy. It can thus happen that, even if the selfish pool
is earning a greater proportion of the total revenue than is indicated
by its share of the total computational power, it is, in fact, earning
revenue at a lesser rate than would be the case if it simply followed
the protocol. This observation makes intuitive sense, since the whole
point of selfish-mine is to put other miners in a position where they
are wasting resources on mining blocks that have no chance of being
included in the long-term \blockchain.

We emphasize that our models, both analytic and simulation, are
idealised. It would be an interesting line for future research to use
network tomography techniques to discover the topology of the actual
Bitcoin network and then employ the analytical and simulation techniques
that we have discussed in this paper to study the effect of propagation
delay on the dynamic evolution of the \blockchain.

\section*{Acknowledgment} \label{sec:acknowledgement}

The authors would like to thank Maria Remerova for some valuable
comments that led to an improvement of the paper. 

\raggedright

\bibliographystyle{IEEEtran}
\bibliography{IEEEabrv,bitcoinFinal_V3}

\appendix \label{appendix}

The pseudo-code presented in Algorithm~\ref{fig:sm} summarises the
actions of a dishonest node.

\begin{algorithm}[b!]

\begin{algorithmic}

{\footnotesize
\caption{Selfish-mine algorithm at a dishonest node~$i$.
\label{fig:sm}}

\Require{Initialise the \textit{blockchain} at dishonest node~$i$.}

\Function{initialise}{}

\State \textit{blockchain} := publicly known blocks

\State \textit{secretExtension} := empty; \textit{race} := \textit{FALSE}

\State mine on the last block in the \textit{blockchain}

\EndFunction

\Statex

\Require{Dishonest node~$i$ attaches a secret \textit{Block} to its
\textit{secretExtension}.}

\Function{secretMine}{\textbf{block} \textit{Block}}

\State append \textit{Block} to \textit{secretExtension}; $n_s := n_s + 1$

\If {\textit{race}}

  \State publish \textit{Block};
         \textit{race} := \textit{FALSE};
         \textit{secretExtension} := empty \label{sm:b}

\ElsIf {$| \textit{secretExtension} | > 5$} \qquad // prevent runaway

\State publish the first unpublished block of \textit{secretExtension} \label{fig:sm:D1}

\EndIf

\State mine on \textit{Block}

\EndFunction

\Statex

\Require{Dishonest node~$i$ attaches a public/published
\textit{Block} to its \textit{blockchain}.}

\Require{The last block on \textit{blockchain} has serial number $n_p$.}

\Require{The last block on \textit{secretExtension} has serial number $n_s$.}

\Function{publicMine}{\textbf{block} \textit{Block}}

\State append \textit{Block} to \textit{blockchain}; $n_p := n_p + 1$

\State $\Delta := n_s - n_p$ \qquad // compute the lead 

\If {$\Delta = -1$}

\If {\textit{race}}

\State \textit{race} := \textit{FALSE}; \label{fig:sm:Dlt0}
       \textit{secretExtension} := empty

\EndIf

\State mine on \textit{Block}
  
\ElsIf {$\Delta = 0$}

\State \textit{race} := \textit{TRUE} \label{fig:sm:D0}

\State publish the last (and only) block of \textit{secretExtension} \label{sm:a}

\State \textit{secretExtension} := empty;
       mine on block $n_s$
  
\ElsIf {$\Delta = 1$}

\If {$|\textit{secretExtension}| = 2$}

\State publish \textit{secretExtension}; \label{fig:sm:D2}
 \textit{secretExtension} := empty

\State mine on block $n_s$

\EndIf
  
\Else \qquad // $\Delta > 1$

\State \mbox{publish the first unpublished block of \textit{secretExtension}}
\label{fig:sm:Dgt1}

\State mine on block $n_s$

\EndIf

\EndFunction
}
\end{algorithmic}

\end{algorithm}

\end{document}